\documentclass[twocolumn,showpacs,preprintnumbers,amsmath,amssymb]{revtex4}
\usepackage{graphicx}% Include figure files
\usepackage{dcolumn}% Align table columns on decimal point
\usepackage{bm}% bold math

\begin{document}

%\preprint{APS/123-QED}

\title{Validity of the scattering length approximation in strongly interacting
Fermi systems}

\author{S. Q. Zhou and D. M. Ceperley }
\affiliation{%
Department of Physics and NCSA, University of Illinois at
Urbana-Champaign, Urbana, Illinois 61801, USA
}%

\author{ Shiwei Zhang}
\affiliation{Department of Physics, College of William and Mary, Williamsburg,
Virginia 23187, USA}

\date{\today}% It is always \today, today,

\begin{abstract}
We investigate the energy spectrum of systems of two, three and
four spin-$\frac{1}{2}$ fermions with short range attractive interactions
both exactly, and within the scattering length approximation.
The formation of molecular bound states and the ferromagnetic
transition of the excited scattering state are examined
systematically as a function of the 2-body scattering length.
Identification of the upper branch (scattering states) is discussed and a general 
approach valid for systems with many particles is given.
We show that an adiabatic ferromagnetic transition occurs, but at a 
critical transition point $k_Fa$ much higher than predicted from previous calculations,
almost all of which use the scattering length approximation. 
In the 4-particle system the discrepancy is a factor of 2. The exact critical
interaction strength calculated in the 4-particle system is consistent with
that reported by experiment. To make
comparisons with the adiabatic transition, we study the quench dynamics of the pairing
instability using the eigenstate wavefunctions.
\end{abstract}
\pacs{Valid PACS appear here}% PACS, the Physics and Astronomy
\maketitle

\section{Introduction}

The control afforded by Feshbach
resonance phenomena in ultracold atomic gases has enabled the
exploration of strongly correlated degenerate Fermi systems. In
a recent study of the possibility of itinerant ferromagnetism
\cite{bloc,ston,zong,duin},
%of delocalized fermions without lattice and band structure,
Jo \textit{et al.} \cite{jo} attempted to observe the physics
behind the Stoner model in an atomic gas of $^{6}$Li atoms.
Evidence for ferromagnetic ordering was seen.
%and saw evidence for ferromagnetic ordering. % in the absence of lattice and band structure.
This experiment has generated a great deal of theoretical
research \cite{lebl, cond, ChangHub2010, troy, triv, reca}. The results have been
debated as to whether a ferromagnetic transition or a strong
correlation effect was seen. Quantitative comparison with
experiment has not been achieved.

A key issue is to find an appropriate model for the experiment.
In the experiment a strong attractive interaction is quickly switched
on. Predictions of the critical ratio of interaction strength to
interatomic spacing for the ferromagnetic transition from mean field theory \cite{ston,lebl},
second order corrections \cite{duin, reca} and QMC calculations
\cite{cond, troy, triv} are on the order of $k_{F}a\sim 1$; 
about two times lower than that from the Jo \textit{et al.}
experiment. In almost all calculations,
a positive interaction \cite{cond, troy} or a Jastrow factor with two-body nodes \cite{triv} is assumed, 
using the scattering length approximation(SLA). Moreover, the various approaches differ in details
of the nature of the transition \cite{ston,duin}.
%Itinerant ferromagnetism is common in
%transition metals. Nevertheless, the microscopic mechanism that
%controls the ferromagnetic transition is not completely
%understood. In the case of electron gas, quantum Monte Carlo
%simulations \cite{zong} suggest that the transition to the
%ferromagnetic state occurs at a critical density about 3 orders
%of magnitude smaller than the prediction of mean-field theory
%\cite{bloc}.REWRITE

%Existing studies  of this experiment assume a pairwise
%interaction using the two-body scattering length approximation
%(SLA) and
The two-body SLA neglects
the low-lying molecular states. %, to which the exact multi-particle scattering state is orthogonal.
The zero-energy $s$-wave scattering length $a$ is defined by
the long distance form of the out-going scattering wave
\begin{equation}
\psi \left(r\rightarrow \infty\right) \propto \frac{\sin\left[
k(r-a) \right]}{kr}  \stackrel{k\rightarrow
0}{\longrightarrow} 1 -\frac{a}{r}.
\end{equation}
For a contact (zero range) potential, $a$ is the radius of the
first wavefunction node: $\psi\left(r=a\right)=0$. The %scattering
%length approximation 
SLA replaces the underlying atomic interaction
by a purely repulsive potential which has the same two-body
scattering length. 

This is analogous to the idea of
pseudo-potentials in electronic structure. A pseudo-potential
can be generated in an atomic calculation to replace the strong
Coulomb potential of the nucleus and the effects of the tightly
bound core electrons by an effective ionic potential acting on
the valence electrons and then used to compute properties of
valence electrons in molecules or solids, since the core states
remain almost unchanged. 
The approach is widely used in electronic structure calculations.
However, it leads to an inaccuate model
if a pseudopotential is used for systems compressed
to high density and the electron cores start to overlap.

Many experiments in cold 
atomic systems are performed near Feshbach resonance where the
scattering length is comparable to
interatomic separation. In this situation, the lower-lying
molecular bound states giving rise to resonance can overlap, 
causing the scattering
states to distort in order to remain orthogonal to the bound
states. With more experimental effort expected in the study of related systems,
precise and reliable comparisons from quantum simulations will be important.
Yet accurate many-body calculations will not be possible without a quantitative 
understanding of the effective interactions and their effect on the different states. 
Even the identification of the scattering state in a dense system requires explanation.

In this article, we quantify this effect by explicitly
including the molecular bound states and treating the interaction exactly.
We consider systems of
two, three and four spin-$\frac{1}{2}$ fermions interacting
through a contact interaction. The energy spectrum as a
function
%of the two-body scattering length
of the two-body interaction strength
is obtained by using an exact
%numerical method on a lattice model and then extrapolate to the
numerical method on a lattice and then extrapolated to the
continuum limit.
We show how the upper branch can be identified for a many-body system.
The properties of the nodal surface of the scattering many-body states are investigated.
% moved to method section:
%The ferromagnetic transition is identified as
%the crossing between the lowest non-magnetic scattering state
%and the fully ferromagnetic state. To investigate the effect of
%using the SLA in multi-particle scattering process, the
%attractive contact interaction is replaced by a zero boundary
%conditions and the resulting critical ferromagnetic transition
%density is estimated.
To compare with the exact solutions, calculations are also made with the SLA,
by replacing the attractive contact interaction with a zero boundary
condition.
%We find that the failure of the
%scattering length approximation is likely responsible for the
%Jo \textit{et al.} results and the calculations employing the SLA.
In both cases an adiabatic ferromagnetic transition is
stabilized. The SLA breaks down for large $a$, leading to
a severe underestimation (by almost a factor of 2) of the
transition point. 
%Our model calculations suggest a value of $k_F a$
%for the ferromagnetic transition consistent to that reported by the 
%experiment.

\section{Method}
\label{sec:method}
We consider a system of two-component
fermions moving in a periodic box with length $L$ to model a
gas of $^{6}$Li atoms with two hyperfine species at non-zero density. All lengths
are expressed in units of $L$ and all energies in units of
$K_{0}=\frac{\hbar^{2}}{2m}(\frac{2\pi}{L})^{2}$. In the case
$a\gg r_{0}$ (where $r_{0}$ is a measure of the effective interaction range), 
the interatomic potential can be modeled as a regularized $\delta$-function:
\begin{equation}
	V(\mathbf{r},\mathbf{r}')=\frac{4\pi\hbar^{2}a}{m}\delta(\mathbf{r}-\mathbf{r}')\frac{\partial}
	{\partial|\mathbf{r}-\mathbf{r}'|}|\mathbf{r}-\mathbf{r}'|
\label{diracdelta}
\end{equation}
where $a$ is the zero-energy scattering length and $m$ is the
mass. %For numerical application,
We solve the Schr\"{o}dinger equation by putting the system on
a lattice with $n$ points in each direction and recover the
continuum limit by extrapolation. We then
approximate the kinetic energy by two different discrete
Laplacian operators \cite{laplacian}: (1) the Hubbard model with nearest neighbor
hopping and (2) a long range hopping model including up to the next nearest
neighbors. We model the bare two-particle interaction by a point contact potential on the grid
\begin{equation}
	V^{\footnotesize{\mbox{grid}}}(\mathbf{r},\mathbf{r}') = -\frac{U}{\Delta^{3}}\delta_{\mathbf{r},\mathbf{r}'},
\label{kronecker}
\end{equation}
where $\Delta = L/n$ is the grid spacing. Here $U>0$ is the
strength of the attractive interaction; on the repulsive side of resonance
$U>U_{\infty}$, we have positive scattering length for unpaired
atoms and the mapping relation between the grid and continuum
is \cite{cast}
\begin{equation}
	\frac{m}{4\pi\hbar^{2}a} = \frac{1}{U_{\infty}}-\frac{1}{U},
\label{mapping}
\end{equation}
where the unitarity point $a\rightarrow\infty$ occurs at
$U^{-1}_{\infty}= (2\pi)^{-3}\int
d^{3}\mathbf{k}(2\epsilon_{\mathbf{k}})^{-1} = \gamma m /(\hbar^{2}\Delta)$. Here $\epsilon_{\mathbf{k}}$
is the single particle dispersion relation and $\gamma$ is a numerical constant defined by the discrete Laplacian. 
For choice (1) above, $\gamma\approx 0.2527$; for choice (2), $\gamma\approx0.2190$. When only nearest neighbor hopping is included, our Hamiltonian is the standard attractive Hubbard model,
but scaled by $1/\Delta^2$. Note our $U$ value scales as $\Delta$, while in the notation 
of the Hubbard model, $U_{\infty}$ is a constant.

In the SLA, $U$ has the opposite sign. In particular, 
when the scattering length $a$ is large, Eq.~(\ref{kronecker}) is replaced by a 
hard-sphere potential with radius $a$.
If $a$ is smaller than the grid spacing, a finite but negative value of $U$ can be used in
the SLA, leading to the repulsive Hubbard model, 
which clearly has a different strongly interacting or 
large $a$ limit \cite{ChangHub2010} from that of Eq.~(\ref{kronecker}).

To determine the eigenvalues and eigenstates, we
start from a set of random trial states
$|\psi_{\alpha}^0\rangle$ where $1\le\alpha\le M$, and evolve the
states $|\psi^{i+1}\rangle = (\mathbf{1}-\tau
\hat{H})|\psi^{i}\rangle$. At each step of the evolution, the state vectors
are properly symmetrized and orthogonalized. As $i\rightarrow
\infty$, the states converge to the lowest $M$ eigenstates of
the Hamiltonian $\hat{H}$ within a given symmetry sector. The
errors are controlled and can be
reduced arbitrarily with increasing the number of grid points
or number of iterations. As discussed below, the computational
cost grows rapidly with system size, but significant reduction
can be achieved by invoking symmetries.

%\begin{equation}
%	\frac{1}{U_{\infty}}\approx\int\frac{d^{3}\mathbf{k}}{(2\pi)^{3}}\frac{1}{\epsilon_{\mathbf{k}}}.
%\end{equation}

\begin{figure}
\scalebox{0.3}[0.3]{\includegraphics[0,0][30cm,22cm]{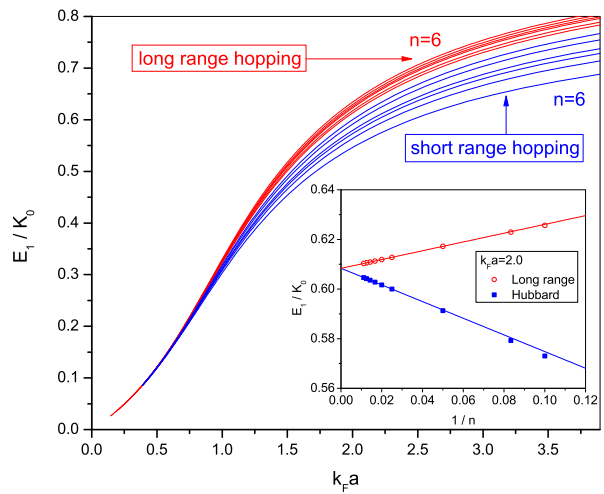}}% Here is how to import EPS art
\caption{\label{fig:tbp} (Color online) The two-body $s$-wave
scattering energy $E_{1}$ as a function of $k_{F}a$
%There is no ferromagnetic transition in the Stoner model of two
%particles in any dimension.
for grid sizes: $n=6,8,10,12,20,40$. The inset shows the
scaling with respect to $1/n$ at $k_{F}a=2.0$ with grid sizes
up to $n=90$. The long-range hopping model converges to the
continuum limit faster than the Hubbard model.}
\end{figure}

%\textit{Test}---
To assess the accuracy of iterative
diagonalization, we test the method on a two-particle problem.
%The following calculations are done in a finite cubic box with side length $L$.
%Note that the lowest excitation of two noninteracting particles
%is $2K_{0}$ because the system has zero total momentum.
The energy of the lowest two-particle scattering state is plotted in Fig~(\ref{fig:tbp}) as a
function of the dimensionless scattering length $k_{F}a$, where
$k_{F}=(3\pi^{2}\rho)^{1/3}$ is the Fermi wave vector and
$\rho$ the particle density. Both discrete representations of
the kinetic energy operator were used and they converge to the same
continuum limit: $n\rightarrow\infty$; the long-range hopping
is found to be less sensitive to the lattice spacing. Solving 
the two-particle problem also enables
us to construct repulsive pseudo-potentials in the SLA by
inverting the 2-particle Schr\"{o}dinger equation.

\section{Two fixed point potentials}
\label{sec:1body}
\begin{figure}
\scalebox{0.8}[0.8]{\includegraphics[0,0][9cm,7.8cm]{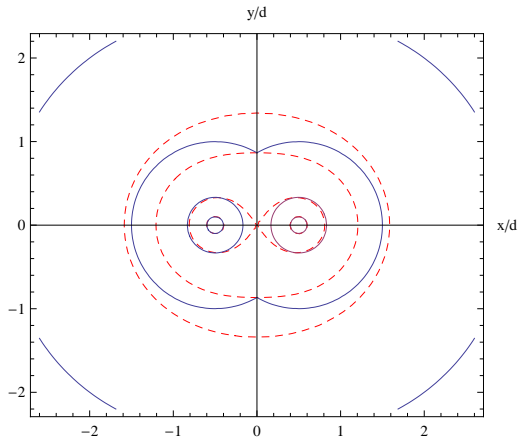}}% Here is how to import EPS art
\caption{\label{fig:tnodes} (Color online) Nodal surface for the scattering wavefunction in a potential
generated by two fixed particles located at $(\pm d/2,0,0)$ in infinite space with $a/d=1/10,1/3,1,5/2$ (expanding outward), where $d$ denotes the distance between the two fixed scatterers.
The solid (blue) lines correspond to the nodes in SLA and the dashed (red) lines to the exact nodes. SLA gives
a reasonable approximation to the nodal surface for $a/d<1$ but the deviation becomes significant for large scattering lengths.}
\end{figure}

The simplest case where the scattering length approximation may
fail is the scattering of a single particle off of two fixed
contact potentials. This problem can be solved exactly in
infinite space \cite{brue}. The nodal surface of the
zero-energy scattering state is given as the solution of:
\begin{equation}
\frac{1}{|\mathbf{r}-\mathbf{R}_{1}|}+\frac{1}{|\mathbf{r}-\mathbf{R}_{2}|}= \frac{1}{a} + \frac{1}{|\mathbf{R}_{1}-\mathbf{R}_{2}|},
\label{exactnodes}
\end{equation}
where $\mathbf{R}_{1}$ and $\mathbf{R}_{2}$ are the location of
the two fixed scatterers. In the SLA, one would model the state
by the ground state with nodes defined by
$|\mathbf{r}-\mathbf{R}_{1}|=a$ and
$|\mathbf{r}-\mathbf{R}_{2}|=a$. The nodal surfaces described by Eq.~(\ref{exactnodes}) 
are shown in Fig (\ref{fig:tnodes}) in comparison with corresponding spheres in SLA.
Clearly the deviation from SLA becomes
significant as $a\sim |\mathbf{R}_{1}-\mathbf{R}_{2}|$. In particular, the spherical surfaces 
in SLA becomes infinitely large at unitarity limit while Eq.~(\ref{exactnodes}) gives
rise to a finite surface. This result suggests that introducing a node in the two-body Jastrow
factor in the form $f(r)\sim (1-a/r)$ is insuffcient to characterize the effective pairwise repulsion 
on the upper-branch \cite{triv}, as further discussed below in Sec.\ref{sec:SLAvsExact}.

\begin{figure}
\scalebox{0.29}[0.29]{\includegraphics[0,0][30cm,23cm]{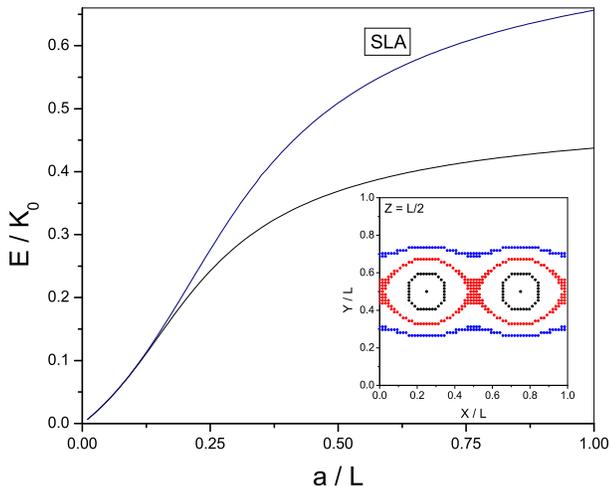}}% Here is how to import EPS art
\caption{\label{fig:twocenter} (Color online) The scattering
energy of a particle moving in the potential generated by two
fixed particles located %on the $x$-axis
at $x=L/4$ and
$x=3L/4$, with $y=L/2$ and $z=L/2$. The SLA is obtained by replacing each potential by a
hard sphere (zero boundary condition) with the same scattering
length. The SLA gives accurate energies in weakly-interacting
limit ($a/L<0.2$) but overestimates the scattering energy for
$a$ comparable with $L$. %This calculation is done on a series
%of finite grids with $N=16,32,64$ and shown on the graph is the
%extrapolation to continuum.
The inset shows the nodal region $(|\psi_{i}|<10^{-4})$ for the scattering states with
$a/L=0.1\mbox{(black)},0.2\mbox{(red)},0.4\mbox{(blue)}$.
The surfaces become noticeably non-spherical for $a/L>0.1$. }
\end{figure}

To study the effect of the SLA at finite density, we solved the
same problem numerically in a finite periodic box. The results
are summarized in Fig~(\ref{fig:twocenter}). It can be seen
from the graph that at large scattering length (i.e. at high
density), the SLA significantly overestimates the scattering
energy for the 3-body problem, i.e. the effect of low-lying
molecular states cannot be ignored. The exact solution achieves
a lower energy by distorting the nodal surfaces away from the
union of two spheres required by the SLA. As we show below,
this also applies to a system of more fermions.

%The $s$-wave
%There is no scattering between fermions with equal spins
%because of antisymmetry; the scattering energy vanishes.
%Nevertheless, the effect of the interaction could be very
%effective for a Fermi gas in non-magnetic states and drives the
%system ferromagnetic.

\section{Four-particle model}
Now consider a minimal model for the ferromagnetic transition:
four spin-$\frac{1}{2}$ atoms in a  cube with periodic boundary
conditions and interacting with a contact potential.
%With periodic boundary conditions, the states with zero total
%momentum are invariant under translations.
The spin polarized state $\Psi(1234) = \psi_{A}(1234) \otimes
\left|\uparrow\uparrow\uparrow\uparrow\right\rangle$ has a
totally antisymmetric spatial part $\psi_{A}(1234)$: for a
contact interaction it is noninteracting with an energy of
$4K_{0}$ in a zero total momentum eigenstate that has 
translational invariance.
%The cubic symmetry allows us
%to assume an appropriate linear combination of eigenstates to
%be invariant under reflections and interchanges of Cartesian
%coordinates. Furthermore, the states of a system of identical
%fermions are totally antisymmetric with respect to particle
%permutations.
On the other hand, there are two linearly independent spin
states with zero total spin $S^{2}=0$ corresponding to
unpolarized states:
\begin{eqnarray}
				\nonumber \chi_{MS} \propto \left|\uparrow\uparrow\downarrow\downarrow\right\rangle
				+\left|\downarrow\downarrow\uparrow\uparrow\right\rangle-\frac{1}{2}\big[\left|\uparrow\downarrow\right\rangle + \left|\downarrow\uparrow\right\rangle\big]\otimes\big[\left|\uparrow\downarrow\right\rangle + \left|\downarrow\uparrow\right\rangle\big],\\
				\nonumber \chi_{MA} \propto \big[\left|\uparrow\downarrow\right\rangle - \left|\downarrow\uparrow\right\rangle\big]\otimes\big[\left|\uparrow\downarrow\right\rangle - \left|\downarrow\uparrow\right\rangle\big].\quad\quad\quad\quad\quad\quad\quad\quad\,\,\;
\end{eqnarray}
The wavefunction is a linear combination of the above two
states $\Psi(1234) =\nonumber \psi_{MA}\otimes \chi_{MS} + \psi_{MS}\otimes
\chi_{MA}$. The symmetries of $\psi_{MA}$ and $\psi_{MS}$ in
coordinate-space are determined by the total antisymmetry of
the complete wavefunction including spins and coordinates.

For a system of four particles on a grid with $n$ points in
each direction, the discretized configuration space has
$n^{12}$ grid points. Translational symmetries along the three
spatial directions reduce the size of the configuration space
by a factor of $n^{3}$. Cubic symmetry of the periodic box
reduces the number of independent wavefunction values by a
factor of $48$ and permutation symmetries give an additional
$24$-fold reduction.
%Thus with the aid of
%reflectional and permutation symmetries, we are able to reduce
%the size of a $9$-dimensional hypercubic by a factor of 1152.
%The actual reduction factor on a grid is slightly smaller
%because some grid points are not affected by all symmetries and
%has an order of magnitude of $10^{3}$ .
We evolve pairs of
non-magnetic states $\{\psi_{MS},\psi_{MA}\}$ within the
reduced domain, and whenever the value of the wavefunctions on
a grid point outside the reduced domain is needed in
off-diagonal projections, the exterior point is mapped back
into the reduced domain by symmetry transformations.
%A neighbor-list can be constructed before the projection process
%to make the mapping efficient. The state vectors are properly
%orthogonalized within the reduced domain at each step
%\begin{equation}
%{\sum_{j}}'\frac{m_{j}}{2}\big[\psi_{MS}(j)\phi_{MS}(j)+\psi_{MA}(j)\phi_{MA}(j)\big] = \delta_{\psi\phi},
%\end{equation}
%where $m_{j}$ denotes the multiplicity of the grid point $j$ in
%the reduced domain with respect to all symmetry
%transformations.

% moved here from introduction -- needs better integration:
The ferromagnetic transition is identified as the crossing
between the lowest singlet scattering state and the fully
ferromagnetic state. To investigate the effect of using the SLA
in multi-particle scattering process, the attractive contact
interaction is replaced by a zero boundary condition and the
resulting critical ferromagnetic density is estimated.

Fig~(\ref{fig:spectrum}) shows the energy spectrum of a
four-particle system for $n=10$ as a function of the coupling
coefficient $U$. %normalized at unitarity.
In this calculation, the lowest 30  states were followed. Note
that we only considered states with the same symmetries as the
ground state, i.e. with even parity with respect to reflection
in $x$, $y$ or $z$. The resulting energy levels can be
classified into three categories by their behavior at strong
coupling: two-molecule states, molecule-atom-atom states and
four-atom scattering states. Level avoiding \cite{land} can be
observed between states belonging to different categories.
%There is a general theorem in quantum mechanics that if the
%Hamiltonian of a finite system contains some parameter and its
%eigenvalues are consequently functions of that parameter,
%energy levels of like symmetry cannot intersect .

\section{Identification of the scattering states}
\label{sec:Idscatter}
The formation of molecular bound states is characterized by the
binding energy diverging linearly as $U\rightarrow \infty$. In
particular, the ground state wavefunction can be approximately
written as $\psi_{0}(13)\psi_{0}(24) -\psi_{0}(14)\psi_{0}(23)$
where $\psi_{0}$ is the two-body bound state, and the ground
state energy is approximately twice the two-particle binding
energy. As seen in Fig. (3) the two-molecule states have an
energy slope ($\partial E / \partial U$) about twice as large
as the molecule-atom-atom states. As $U\rightarrow \infty$
molecules become tightly bound; their energy spacings can be
understood in terms of colliding molecules. For a lattice
model, in contrast to a continuum model, the greater the
internal binding energy, the greater the total mass of the
molecule \cite{matt}.

\begin{figure}
\scalebox{0.3}[0.3]{\includegraphics[0,0][30cm,23.8cm]{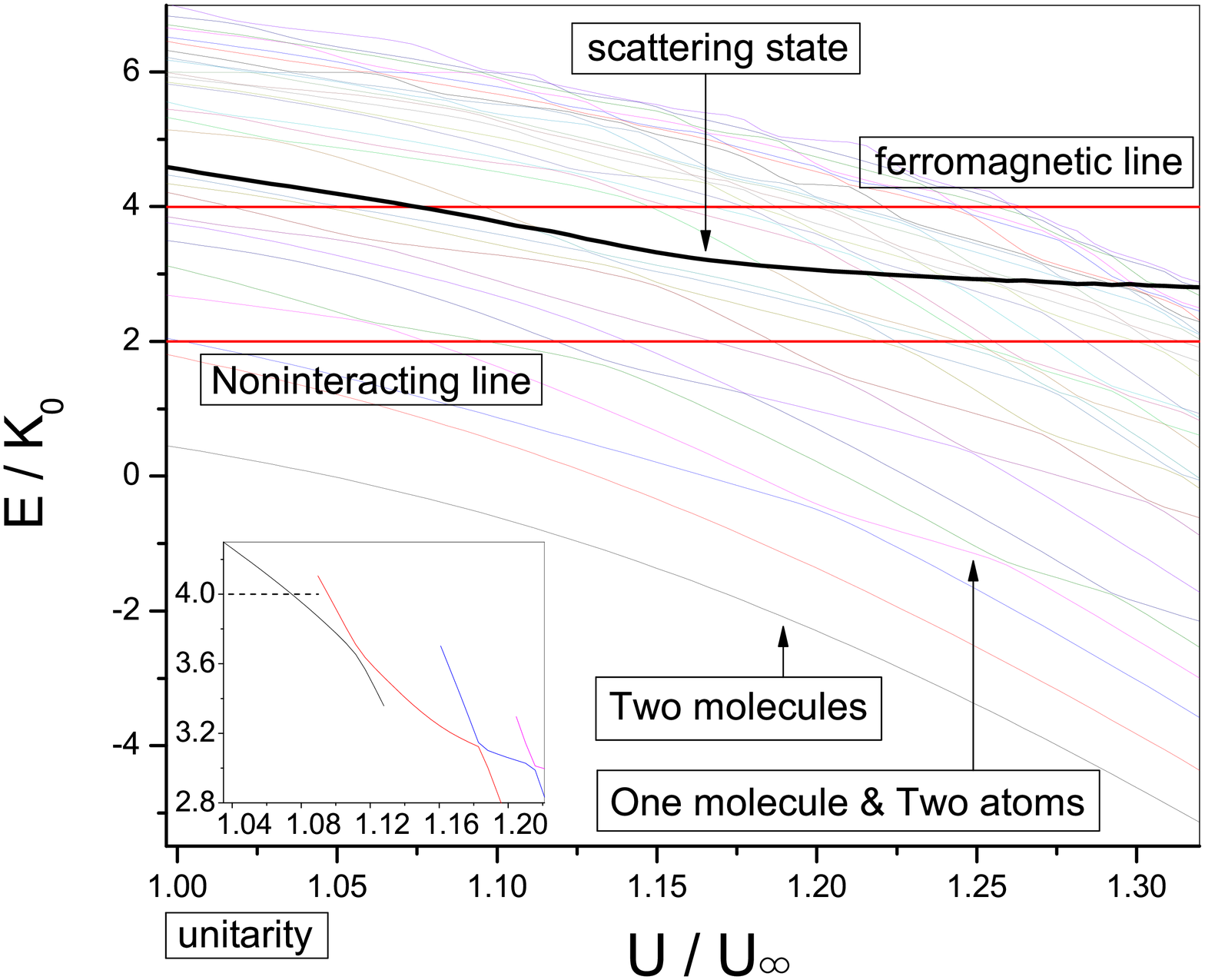}}% Here is how to import EPS art
\caption{\label{fig:spectrum} (Color online) The energy
spectrum of the lowest 30 $s$-states of four fermions with
contact interactions on a 3D grid with $n=10$. The energy
levels can be classified into two-molecule states,
molecule-atom-atom states and four-atom states. The ferromagnetic transition can be identified as
the lowest scattering state (heavy dark line) crossing the horizontal
ferromagnetic line around $U/U_{\infty}\approx 1.07$. The inset shows an enlargement of the lowest scattering
state and the associated avoided crossings.}
\end{figure}

The scattering state of strongly repulsive atoms is an excited branch and all cold atom experiments performed in this regime are metastable. In Jo \textit{et al.} experiment, the magnetic field ramp ($\sim 4.5$ms) is much slower compared to the characteristic time
scale of atomic collisions $\hbar/k_{B}T_{F}\sim \mu$s, and marches toward the resonance from the repulsive side $a\gtrsim 0$. At low density or in the weakly interacting
regime, the four-atom scattering state approaches the noninteracting line $2K_{0}$ and the 
SLA is an accurate approximation. But there are some difficulties in defining the 
scattering state at high density or in the strongly interacting regime because of the level avoiding phenomena. As shown in the inset of Fig~(\ref{fig:spectrum}), if the coupling coefficient $U$ is
tuned toward the resonance $U_{\infty}$, it is energetically more favorable to jump through the 
successive avoided crossings. Thus, the change in the scattering
energy due to an adiabatic tuning of the interaction can then be determined by following the excited branch curve. It is drawn in bold in Fig~(\ref{fig:spectrum}).

There is another way to identify the upper branch (scattering states) quantitatively by using the momentum distribution and the pairing order. First, consider the wavefunctions
for the relative motion of two particles interacting through a large scattering length 
of Eq.~(\ref{diracdelta}).
The zero-energy scattering state in coordinate space $\psi(\mathbf{r})\propto r^{-1}-a^{-1}$
takes the form $\psi(\mathbf{k})\propto 4\pi k^{-2} - (2\pi)^{3}a^{-1}\delta(\mathbf{k})$
in momentum space and diverges at $\mathbf{k}=0$. By contrast, the bound state $\psi(\mathbf{r})\propto r^{-1}e^{-r/a}$
takes the form $\psi(\mathbf{k})\propto 4\pi (k^{2}+a^{-2})^{-1}$
in momentum space and remains finite at $\mathbf{k}=0$. The momentum distribution $n(\mathbf{k})$ 
for scattering states is different from bound states at $\mathbf{k}=0$: scattering states have a larger
fraction of particle occupation at $\mathbf{k}=0$. 

We also consider the pairing order defined by:
\begin{equation}
g_{2}\equiv n \Big\langle
\sum_{i<j}\delta_{\mathbf{r}_{i},\mathbf{r}_{j}}
\Big\rangle_{\alpha} 
\label{g2def}
\end{equation}
for each state $|\psi_\alpha\rangle$.
The quantity $g_2$ measures double occupancy, and is related to the energy slope: 
\begin{equation}
\frac{\partial E_{\alpha}}{\partial U} = \Big\langle \frac{\partial
\hat{H}}{\partial U}  \Big\rangle_{\alpha} =
-\frac{1}{\Delta^{2}L}g_{2}.
\label{dEdU}
\end{equation}
For the scattering state, double occupancy decreases monotonically as the interaction 
strength is increased (see e.g.~Ref.~ \cite{ChangHub2010}). 
Thus the scattering state in each lattice system is 
characterized by a vanishing energy slope as $U\rightarrow \infty$,
\begin{equation}
g_2 \rightarrow 0, \quad \frac{\partial E_{\alpha}}{\partial U} \rightarrow 0,
\label{doublecount}
\end{equation}
as can be seen in Fig~(\ref{fig:spectrum}).
The pairing density is also related to the tail of  the momentum distribution, 
which describes the short range physics.
At large $k$,  the momentum distribution takes the form $n(\mathbf{k}) \rightarrow C/k^{4}$,
%for both the bound states and scattering states, 
where the coefficient $C$ is called the \textit{contact} 
in the Tan relations \cite{tan}. In the continuum limit
$\Delta\ll a$, the contact $C$ 
can be related to the energy slope, and hence $g_2$, 
%number of double occupancy on a grid 
through the adiabatic sweep theorem
$\frac{dE}{da^{-1}} = -\frac{\hbar^{2}}{4\pi m}C$, which in our system yields:
\begin{equation}
g_{2}= L\Big[\gamma - \frac{\Delta}{4\pi a}\Big]^{2}\,C,
\label{g2contact}
\end{equation}
where 
%$C$ is units of $L^{-1}$ and 
$\gamma$ is the numerical constant appearing in the definition of $U_{\infty}$.
For bound states this gives a finite $g_2$ and a finite energy slope.
% which, in the BEC limit, scales as $1/\Delta^3$ with grid spacing.

\begin{figure}
\scalebox{0.3}[0.3]{\includegraphics[0,0][30cm,23.6cm]{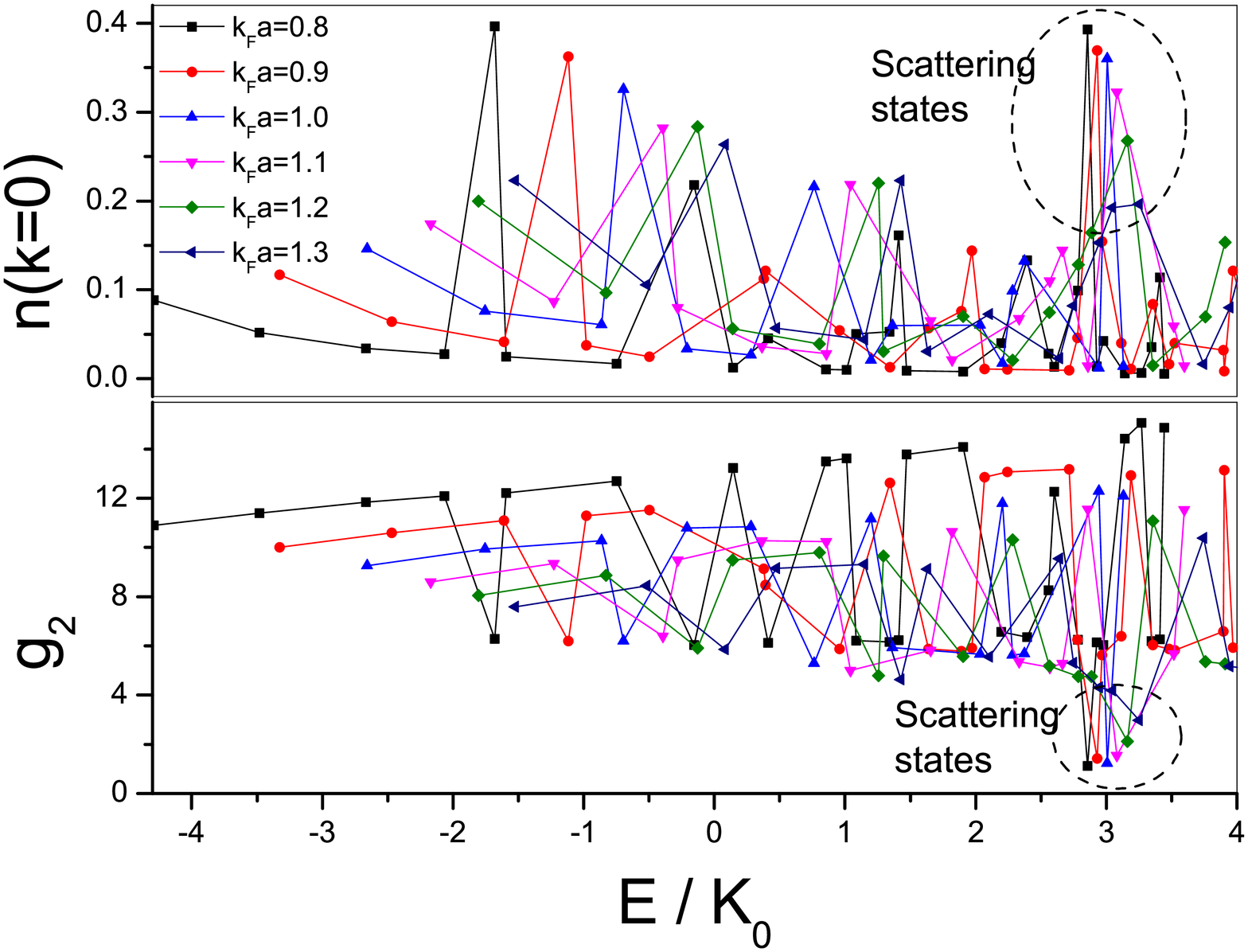}}% Here is how to import EPS art
\caption{\label{fig:g2v} (Color online) The momentum distribution $n(\mathbf{k}=0)$ and the pairing parameter $g_{2}$ of the lowest 30 $s$-states of four fermions with
contact interactions on a 3D grid with $n=10$ versus energy. Scattering states have, by definition, $E/K_{0}>2$ and can be identified by the large magnitude of $n(\mathbf{k}=0)$ and small magnitude of $g_{2}$ compared to the other states. The peak structure
at the scattering state diminishes as the interaction parameter $k_{F}a$ increases.}
\end{figure}

Thus, in addition to the momentum distribution at $\mathbf{k}=0$, we can identify the scattering state from
the other states by the magnitude of $g_{2}$: scattering states have
a smaller fraction of double occupation
$\mathbf{r}_{i}=\mathbf{r}_{j}$. Note that the contact $C$ measures
the local density of pairs \cite{braa}. The momentum distribution $n(\mathbf{k}=0)$ and the pair parameter $g_{2}$ are plotted in Fig~(\ref{fig:g2v}) as functions of the energy for $k_{F}a=0.8\sim 1.3$. Scattering states are, by definition, in the range $E/K_{0}>2$ and can be identified by the peaks of $n(\mathbf{k}=0)$ and low values of $g_{2}$. 

\section{\label{sec:SLAvsExact}Comparison of the SLA and exact results}
\begin{figure}
\scalebox{0.3}[0.3]{\includegraphics[0,0][30cm,23.6cm]{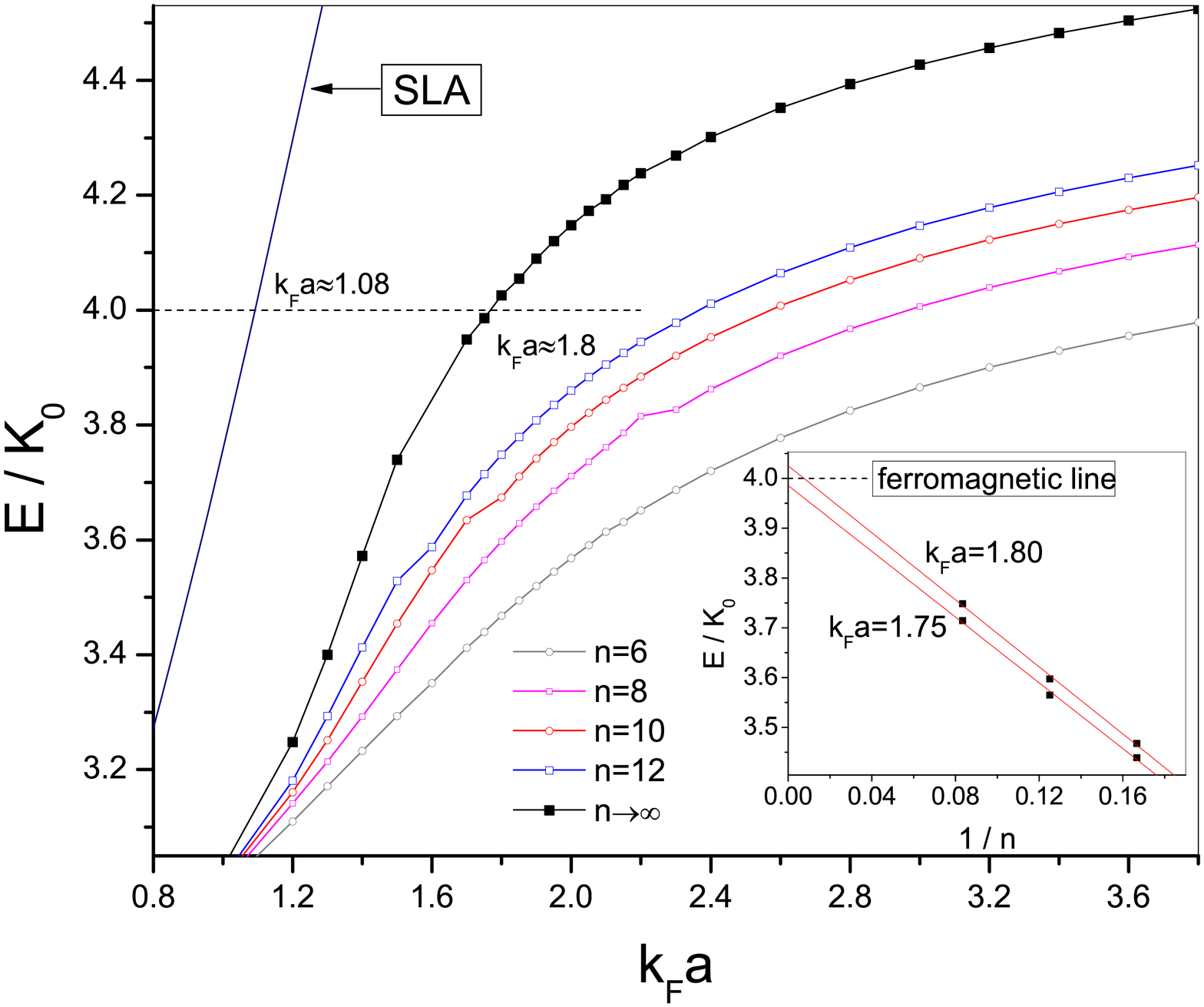}}% Here is how to import EPS art
\caption{\label{fig:fourbody} (Color online) The four particle
scattering energy as a function of the scattering length. The
energy of the ferromagnetic state is shown as the dashed
horizontal line. The extrapolation to continuum is performed
based on calculations on grids with $n=6,8,10,12$, which exhibits 
the $1^{\mbox{st}}$-order linear scaling $1/N$ to high accuracy. The
transition to the ferromagnetic phase occurs at $k_{F}a\approx
1.8$ while the SLA  gives the transition at $k_{F}a\approx
1.08$. The inset shows the scaling with respect to $1/n$ near the transition point $k_{F}a=1.75$ and $1.8$.}
\end{figure}

The ferromagnetic transition
in the four-atom system occurs when the scattering state
energy equals the noninteracting energy, $4K_{0}$. For a $n=10$
grid, the transition occurs at $U/U_{\infty}\approx 1.07$.
Shown in Fig~(\ref{fig:fourbody}) is the energy of the
four-particle unpolarized scattering state as a function of the
scattering length $k_{F}a$ on grids with $n=6,8,10,12$ and
their extrapolation to the continuum limit, $n=\infty$. Avoided
crossings between levels appear as kinks. The excited
scattering state from the  solution of the four-particle
problem crosses the ferromagnetic line at $k_{F}a\approx 1.8$,
which is in remarkable agreement with the reported experimental value of $k_{F}a=1.9\pm 0.2$
\cite{jo}. 

Also shown is the scattering energy using the SLA;
this gives a ferromagnetic transition at $k_{F}a\approx 1.08\sim 1.09$ for grid sizes $n=8,10,12$,
consistent with previous calculations \cite{ston, duin,
cond, troy, triv, reca}. The earliest Fixed-node diffusion Monte Carlo calculations employed
the repulsive P\"{o}schl-Teller potential ($k_{F}a\approx 0.86$) \cite{cond}, hard spheres or repulsive
soft spheres ($k_{F}a\approx 0.82$) \cite{troy} and included backflow effects ($k_{F}a\approx 0.96$) \cite{triv}. For attractive
interactions modeled by spherical square wells or attractive P\"{o}schl-Teller potential, either variational
Monte Carlo ($k_{F}a\approx 0.86$) \cite{troy} or fixed-node diffusion Monte Carlo \cite{triv} ($k_{F}a\approx 0.89$) calculates the upper-branch metastable state by
imposing a nodal condition in the many-body wave function. The nodal condition ensures
that the calculation consist of unbound
fermionic atoms and no dimers or other bound molecules, 
by introducing a Jastrow factor in the form of the scattering solution
of the attractive potential corresponding to positive energy.
As shown in Sec.~\ref{sec:1body}, the nodal structure obtained this way deviates significantly 
from the true nodes in the upper branch. This explains why  
all these calculations gave results similar to those from repulsive potentials, and all of them reproduced
the predicted behavior of the mean field theory and second order corrections. 

The discrepancy between
the critical values of $k_{F}a$ reflects the limitations of perturbation theory in the regime
of strong coupling. Compared to repulsive potentials, using Jastow factor with nodes and including backflow effects
for attractive potentials improves 
%the agreement with experiment 
the result by making nontrivial modifications to the nodal structure, but still gives answers not qualitatively different from the repulsive potential, and fails to reveal the inadequacy of the SLA.

These observations suggest that lower-lying molecular states are
responsible for delaying the formation of the ferromagnetic
phase.  However, calculations with more than 4 atoms are needed
to determine finite size effects. Such calculations are not
feasible with the current method but might be possible with
stochastic methods.

\section{Dynamics of four-particle model}
Because of the limited lifetime of the strongly interacting
gas, however, the magnetic field ramp in experiment is not
adiabatic and can lead to different explanations\cite{pekk,zhai}.
A recent work \cite{cond2} takes into account
the effect of atom loss by including a fictitious three-body term in the
effective Hamiltonian of the Fermi gas and found that the critical
interaction strength required to stabilize the ferromagnetic state
increases significantly. A full $T$-matrix analysis \cite{pekk} suggests that the pairing 
instability can prevail over the ferromagntic instability
and the experimentally measured atom loss rate can be qualitatively explained
in terms of the growth rate of the pairing order parameter after a quench.

\begin{figure}
\scalebox{0.3}[0.3]{\includegraphics[0,0][30cm,23.6cm]{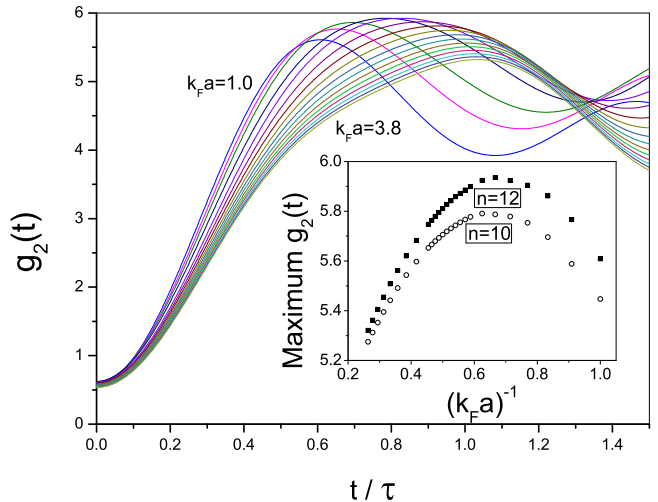}}% Here is how to import EPS art
\caption{\label{fig:evolution} (Color online) The first oscillation period in the evolution of $g_{2}(t)$ after a quench from a non-interacting state to an interaction strength $k_{F}a$, where $\tau = \hbar/\epsilon_{F}$. The inset shows the nonmonotonic behavior of the maximum value of $g_{2}(t)$ in the first oscillation period as a function of the final interaction strength $k_{F}a$. This calculation in done for four particles on a grid with $n=12$.}
\end{figure}

Thus, it is an interesting problem to study the dynamics of the 
pairing instability after a quench using the wave functions obtained  in this
work. Since the contact $C$ is identified as the integral over space of the expectation value of a local
operator that measures the density of pairs \cite{braa}, we characterize the pairing instability by the 
count of double occupancy $g_{2}$ in Eq (\ref{g2contact}).
%The adiabatic change
%of $g_{2}$ in the four-body ground state under the tuning of the scattering length from the tight-binding 
%limit $a\rightarrow 0^{+}$ toward the BEC side of the resonance is shown in Fig (\ref{fig:g2adiabatic}). 
%The asymptotic relation $g_{2}\sim (k_{F}a)^{-1}$ can be observed in the BEC limit \cite{hulet}.
%\begin{figure}
%\scalebox{0.3}[0.3]{\includegraphics[0,0][30cm,23.6cm]{Fig06.eps}}% Here is how to import EPS art
%\caption{\label{fig:g2adiabatic} The expectation value of $g_{2}$ in the four-body unpolarized ground state as a function 
%of the scattering length from the tight-binding 
%limit $a\rightarrow 0^{+}$ toward the BEC side of the resonance. This calculation in done on a grid with $n=12$.}
%\end{figure}
To study the dynamics of the pair formation, we choose the initial state to be the
unpolarized four-particle ground-state in the noninteracting limit and expanded in the basis of the lowest $16$ eigenstates with the final interaction after the quench. The time evolution is then evaluated using the eigenstate expansion $|\psi(t)\rangle = \sum_{m}c_{m}e^{-\frac{i}{\hbar}E_{m}t}|\phi_{m}\rangle$. 
The pairing density $g_{2}(t)=\langle \psi(t)|g_{2}|\psi(t) \rangle$ is used to characterize the pair formation and the atom loss into molecules. The evolution diagram of $g_{2}(t)$ is shown in Fig (\ref{fig:evolution}). The nonmonotonic dependence of the maximum value of $g_{2}(t)$ in the first oscillation period on the final scattering length $k_{F}a$ after the quench is in qualitative agreement with experiment and the full $T$-matrix theory \cite{pekk}. 

\section{Summary}
In summary, we have assessed the accuracy of the scattering length approximation
%for multi-particle scattering process
at high density or strong interaction % <-- talking to cold atoms
$k_{F}a \gtrsim 1$. It is demonstrated that if molecular states
mix with excitations, non-magnetic states are stabilized. 
Identification of the upper branch in many-body calculations is discussed.
The corresponding nodal structures of the states are examined.
The calculated critical interaction strength $k_{F}a$ for ferromagnetic transition is shown to be 
underestimated by a factor of two under the scattering length approximation.
Although we solved the problem
only for 4 particles, this minimal model suffices to show that ignoring the molecular states 
with the scattering length approximation leads to 
inaccurate results in the strongly interacting regime. Hence it leads to severe errors in many-body calculations.
That we get very good agreement with experimental estimates is encouraging but could be a result of cancellation of errors between the 4 particle system and the thermodynamic limit.
%Under the assumption
%of an adiabatic tuning of the interaction parameter, the
%critical value of $k_{F}a$ for the ferromagnetic transition of a
%system of four fermions is found to agree with experiment when the
%interaction is treated properly. 
We investigated the dynamics of pair formation.
Non-monotic behavior of the pairing parameter $\sum_{i<j}\delta_{\mathbf{r}_{i},\mathbf{r}_{j}}$ 
is observed as a function of the final interaction strength $k_{F}a$ after a quench.

\textit{Acknowledgments}---This work has been supported
with funds from the DARPA OLE Program and ARO (Grant
no.~56693PH), computer time at NCSA. This work was initiated at
the Kavli Institute of Theoretical Physics in Beijing and Santa
Barbara.

\newpage %Just because of unusual number of tables stacked at end

\begin{thebibliography}{9}
\bibitem{laplacian} A finite-difference expression for the Laplacian operator is $-\nabla^{2}\psi_{i}=\Delta^{-2}\sum_{j=-n/2}^{n/2}c_{j}\psi_{i+j}$. The coefficients $c_{j}$ are obtained through the Fourier expansion of the eigenvalues of the Laplacian $k^{2}=c_{0}+2\sum_{j=1}^{n/2}c_{j}\cos(kj)$ and are given
in Table I of  Ref.\cite{chel}. This expression also gives the single particle dispersion relation $\epsilon_{k}$ in one dimension.
\end{thebibliography}

\end{document}